\documentclass[aps,prl,floatfix,twocolumn]{revtex4}
\usepackage{graphics}
\usepackage{dcolumn}
\usepackage{epsfig}

\newcommand{\simlt}
      {\ifmmode       { \raisebox{-.8em}{$<$}\atop\sim}
         \else        {$\raisebox{-.8em}{$<$}\atop\sim$}
      \fi}

\begin{document}

\title
{Landau Levels and Quantum Hall Effect in Graphene Superlattices}
\author{Cheol-Hwan Park$^{1,2}$}
\author{Young-Woo Son$^{3}$}
\author{Li Yang$^{1,2}$}
\author{Marvin L. Cohen$^{1,2}$}
\author{Steven G. Louie$^{1,2}$}
\email{sglouie@berkeley.edu}
\affiliation{$^1$Department of Physics, University of California,
Berkeley, California 94720 USA\\
$^2$Materials Sciences Division, Lawrence Berkeley National
Laboratory, Berkeley, California 94720 USA\\
$^3$School of Computational Sciences, Korea Institute
for Advanced Study, Seoul 130-722, Korea}
\date{\today}
\begin{abstract}
We show that, when graphene is subjected to an appropriate one-dimensional
external periodic potential,
additional branches of massless fermions
are generated with nearly the same electron-hole crossing energy
as that at the original Dirac point of graphene.
Because of these new zero-energy branches, the
Landau levels at charge neutral filling
becomes $4(2N+1)$-fold degenerate
(with {\it N}=0,\,1,\,2,\,..., tunable by the potential strength and
periodicity) with
the corresponding Hall conductivity $\sigma_{xy}$ showing
a step of size $4(2N+1)\,e^2/h$.
These theoretical findings are robust against variations in the details of the external
potential and provide measurable signatures
of the unusual electronic structure of graphene superlattices.
\end{abstract}
\maketitle

The physical properties of
graphene~\cite{novoselov:2005Nat_Graphene_QHE,zhang:2005Nat_Graphene_QHE,berger:2006Graphene_epitaxial}
are currently
among the most actively investigated topics in condensed matter physics.
Graphene has the unique feature that the low-energy charge carriers are
well described by the two-dimensional (2D) massless Dirac equation, used for massless neutrinos,
rather than by the Schr\"{o}dinger
equation~\cite{novoselov:2005Nat_Graphene_QHE,zhang:2005Nat_Graphene_QHE}.
Moreover, graphene is considered to be a promising candidate
for electronics and spintronics
applications~\cite{wu:026801}.

It has been shown that, because of their gapless
energy spectrum and chiral nature, the
charge carriers in graphene are not hindered by
a slowly varying electrostatic potential barrier at normal
incidence~\cite{katsnelson:2006NatPhys_Graphene_Klein},
analogous to the Klein tunneling effect predicted in
high-energy physics. Direct evidences of Klein
tunneling through a single barrier in graphene~\cite{katsnelson:2006NatPhys_Graphene_Klein}
have been observed in recent experiments~\cite{stander:026807,young:2009NatPhys}.

Application of multiple barriers or periodic potentials,
either electrostatic~\cite{bai:2007PRB_Graphene_SL,park:2008NatPhys_GSL,barbier:115446,park:2008NL_Supercollimation,park:126804}
or magnetic~\cite{masir:235443,masir:035409,dellanna:045420,ghosh:arxiv},
to graphene has been shown to modulate its electronic structure
in unique ways
and lead to fascinating new phenomena and possible applications.
Periodic arrays of
corrugations~\cite{guinea:075422,wehling:EPL2008_graphene_corrugation,isacsson:035423}
have also been proposed as graphene superlattices (GSs).

Experimentally, different classes of GSs
have been fabricated recently.
Patterns with periodicity as small as 5~nm have been imprinted
on graphene through electron-beam induced deposition of
adsorbates~\cite{meyer:123110}.
Epitaxially grown graphene on the (0001)
surface of ruthenium~\cite{marchini:2007PRB_Graphene_Ru,vazquez:2008PRL_Graphene_SL,
pan:2007condmat_Graphene_SL,sutter:2008NatMat,martoccia:126102} and that on the (111) surface of
iridium~\cite{coraux:2008NL,ndiaye:2008NJP,pletikosic:056808}
also show superlattice patterns with $\sim$3~nm lattice period.
The amplitude of the periodic potential applied to graphene
in these surface systems has been estimated to be
in the range of a few tenths of an electron volt~\cite{vazquez:2008PRL_Graphene_SL}.
Fabrication of periodically patterned gate electrodes
is another possible way of making GSs
with periodicity close to or larger than~$\sim$20~nm.

The quantum Hall plateaus in graphene
take on the unusual values of $4(l+1/2)\,e^2/h$
where $l$ is a non-negative integer~\cite{PhysRevLett.95.146801}.
The factor 4 comes from the spin and valley degeneracies.
In bilayer graphene,
the quantum Hall plateaus are at $4l\,e^2/h$ with $l$ a positive
integer~\cite{mccann:086805}. These unconventional
quantum Hall effects have been experimentally
verified~\cite{novoselov:2005Nat_Graphene_QHE,zhang:2005Nat_Graphene_QHE,novoselov:2006NatPhys},
providing evidences for 2D massless
particles in graphene and massive particles in bilayer graphene.

In this paper, we investigate the LLs
and the quantum Hall effect in GSs formed by the application of
a one-dimensional (1D) electrostatic periodic potential
and show that they exhibit additional unusual properties.
We find that, for a range of potential shapes and parameters,
new branches of massless fermions are
generated with electron-hole crossing energy the same as that
at the original Dirac point of pristine graphene.
These additional massless fermions affect the LLs qualitatively.
In particular, the LLs with energy corresponding to the Fermi energy at
charge neutrality (i.\,e.\,, zero carrier density) become
$4(2N+1)$-fold degenerate ($N=0,\,1,\,2,\,...$), depending on
the strength and the spatial period of the potential
(pristine graphene corresponds to $N=0$).
Accordingly, when sweeping the carrier density from
electron-like to hole-like, the quantum Hall conductivity
in such a GS is predicted to show an unconventional step size of $4(2N+1)\,e^2/h$
that may be tuned by adjusting the external periodic potential.

In our study, the electronic structure of the GSs is evaluated using the methods
developed in Ref.~\onlinecite{park:2008NatPhys_GSL};
we evaluate the bandstructure of the GS numerically by solving the 2D massless
Dirac equation with the external periodic potential included using a planewave basis.
Similarly, to obtain the LLs,
the eigenstates of the GSs under an external perpendicular
magnetic field are expanded with planewaves.
We work in a Landau gauge with the vector potential depending on the position
coordinate along the direction of the periodicity of the GS, and a zigzag
form for the vector potential with a very large artificial
periodicity (large compared to the GS periodicity)
is employed to mimic a constant magnetic field
near the origin in position space~\cite{note:degeneracy}.
{We have checked that the LLs are converged in energy to within less
than 1~\% with respect to the size of the supercell for the vector potential
and the kinetic energy cutoff for the planewaves. The size of the largest
supercell and that of the smallest sampling distance in real space
used are 400 and 0.05 in units of a single unit cell, respectively.}

\begin{figure*}
\includegraphics[width=1.6\columnwidth]{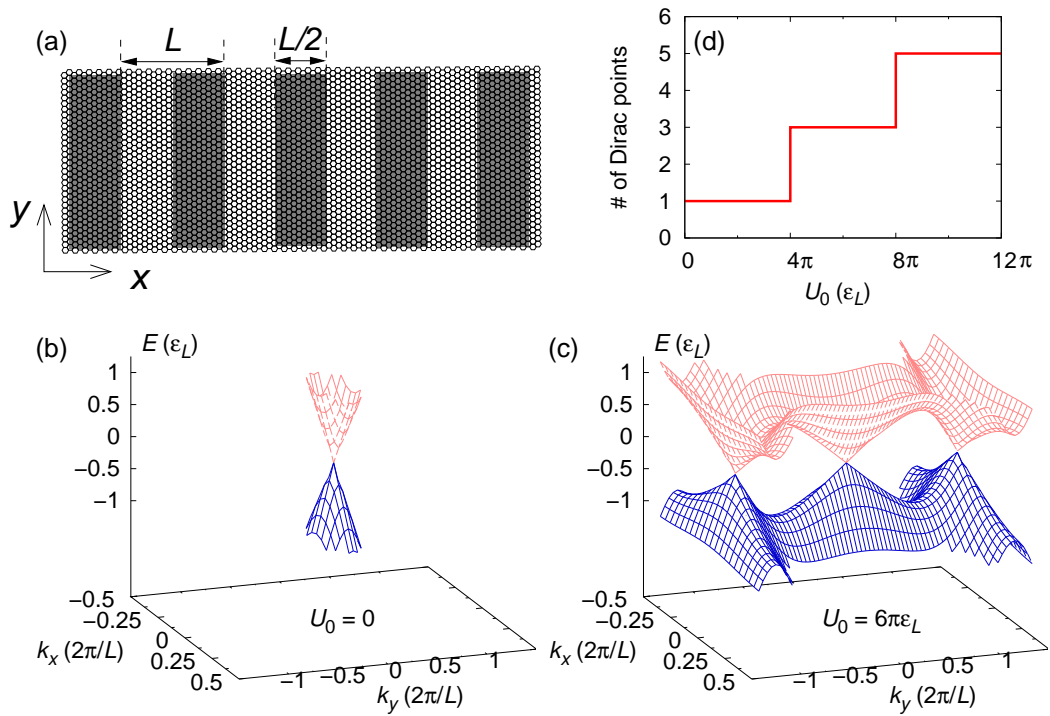}
\caption{(color online) (a) Schematic diagram of a Kronig-Penney type of potential
applied to graphene with strength $U_0/2$ inside the gray regions
and $-U_0/2$ outside with lattice period {\it L} and barrier width
{\it L}/2. (b) Electron energy in units of $\varepsilon_L$
($\equiv\hbar v_0/L$; for example, if $L=20$~nm, $\varepsilon_L=33$~meV)
versus wavevector near the Dirac point in pristine graphene.
(c) The same quantity as in (b) for a GS with $U_0=6\pi\varepsilon_L$.
(d) Number of Dirac points (not including spin and valley degeneracies)
in a GS versus $U_0$.}
\label{Fig1}
\end{figure*}

Figure~\ref{Fig1}(a) shows a GS formed by
a Kronig-Penney type of electrostatic potential periodic along the {\it x} direction,
with lattice parameter $L$ and barrier width $L/2$. Remarkably, unlike
that in graphene [Fig.~\ref{Fig1}(b)], the bandstructure
in a GS [Fig.~\ref{Fig1}(c)] can have, depending on the potential barrier height $U_0$,
more than one Dirac point with $k_x=0$
having exactly the same electron-hole crossing energy~\cite{note:diff_new_gen}.
As Fig.~\ref{Fig1}(c) shows, the number of Dirac points for this type of GSs increases by two
(without considering the spin and valley degrees of freedom)
whenever the potential amplitude exceeds a value of
\begin{equation}
U_0^N=4\pi N\,\hbar v_0/L
\label{eq:U0}
\end{equation}
with $N$ a positive integer. The value of the potential barrier given
in Eq.~(\ref{eq:U0}) corresponds to
special GSs in which the group velocity
along the $k_y$ direction vanishes
for charge carriers whose wavevector is near
the original Dirac cone [e.\,g.\,, the Dirac cone
at the center in Fig.~\ref{Fig1}(c)]~\cite{note:vperp,note:validity_Dirac}.
All the findings in this study apply in general to GSs made from a
periodic potential which has both even and odd symmetries, like a sinusoidal type
of potential.
The results for GSs whose odd or even symmetry is broken are discussed
in Ref.~\onlinecite{park:unp_chiral}.

\begin{figure}
\includegraphics[width=1.0\columnwidth]{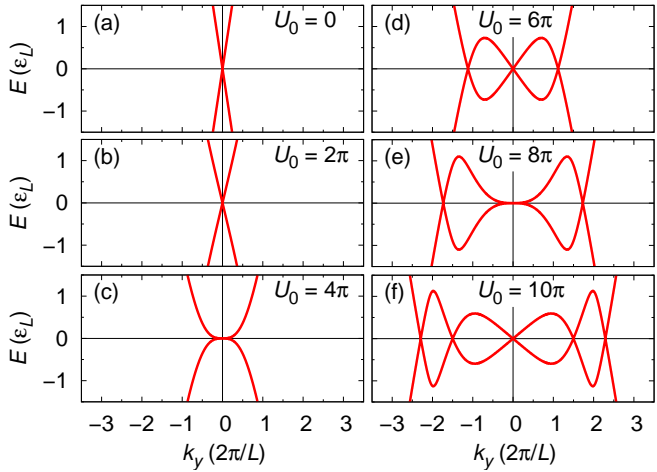}
\caption{(color online) Electron energy (in units of $\varepsilon_L=\hbar v_0/L$)
versus $k_y$ with $k_x=0$ in GSs shown in Fig.~\ref{Fig1} for several different values of barrier
height $U_0$ (specified in each panel in units of $\varepsilon_L$).}
\label{Fig2}
\end{figure}

Figure~\ref{Fig2} shows the evolution of the energy of the electronic states
with $k_x=0$ for a GS depicted in Fig.~\ref{Fig1} for several different values of $U_0$.
As stated above, the group velocity along the $k_y$ direction becomes zero
near $k_y=0$ when the barrier height is given by Eq.~(\ref{eq:U0})
[Figs.~\ref{Fig2}(c) and~\ref{Fig2}(e)].
When $U_0$ has a value between those specific values,
the position of the additional new Dirac points move away from the $k_y=0$
point along the $k_y$ direction with increasing $U_0$.
The complex behavior of the zero-energy Dirac cones revealed by our numerical
calculations cannot be derived using perturbation theory~\cite{park:126804} because
$k_y$ is not small compared to the superlattice reciprocal lattice spacing $2\pi/L$.
Moreover, the pseudospin character of these additional massless fermions [e.\,g.\,,
the left and the right Dirac cones (not the center one) in Fig.~\ref{Fig1}(c)]
are different from that of the original massless Dirac fermions. For example,
backscattering amplitude due to a slowly varying potential
within one of the new cones does not vanish~\cite{park:unp_chiral}.

\begin{figure}
\includegraphics[width=0.8\columnwidth]{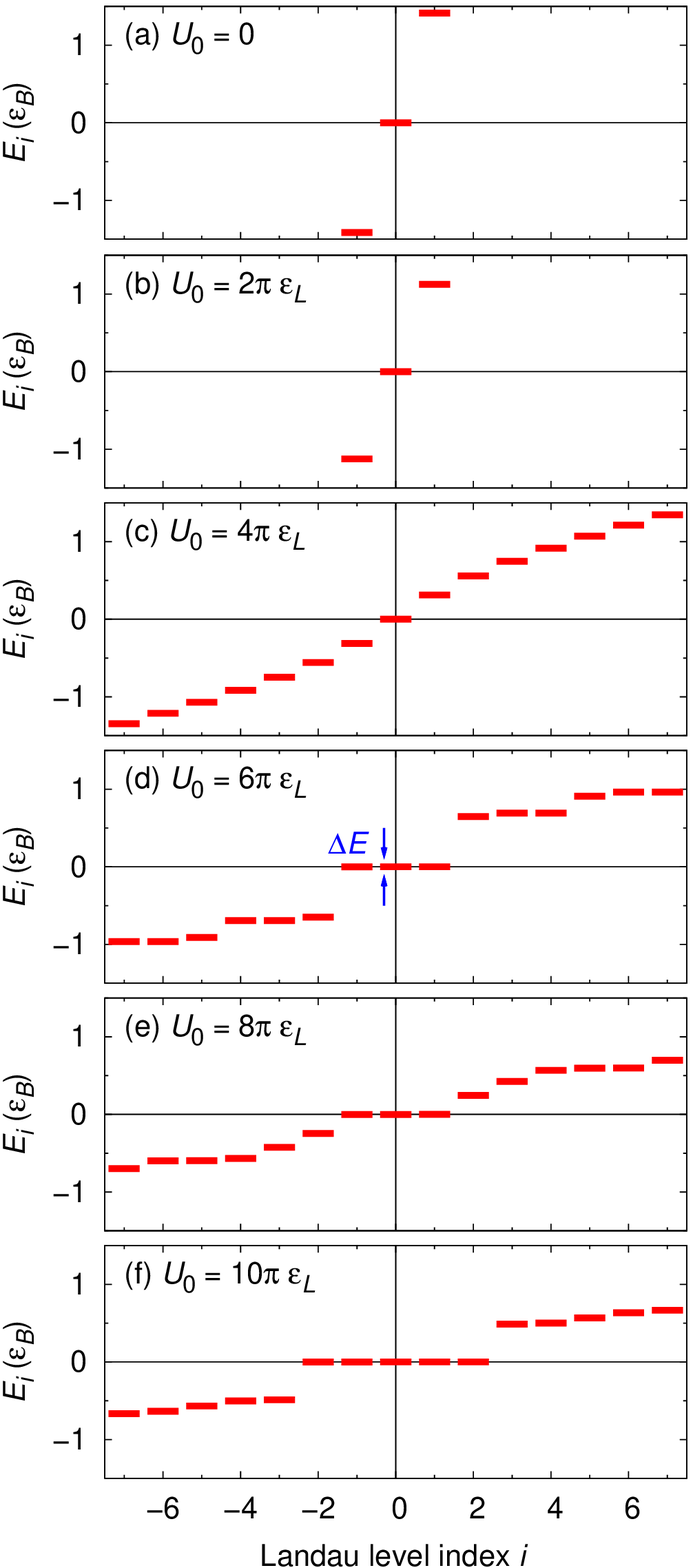}
\caption{(color online) Landau level energy $E_i$ (in units of
$\varepsilon_B\equiv\hbar v_0/l_B$ with $l_B=\sqrt{\hbar c/eB}$)
versus the Landau level index $i$ ($i=0,\,\pm1,\,\pm2,\,...\,$)
in GSs formed with a 1D Kronig-Penney potential
for several different values of barrier height $U_0$, with lattice period $L=0.5l_B$.
The LLs now have a finite width $\Delta E$ (shown not to scale and
exaggerated in the figure)
arising from the $k_y$ dependence of the energy of the electronic states
in a perpendicular magnetic field~\cite{PhysRevLett.62.1177}.
Note the 3-fold and the 5-fold degeneracies around $E_i=0$ in (d) and (f), respectively.
(If the spin and valley degeneracies are considered, those become
12-fold and 20-fold, respectively.)}
\label{Fig3}
\end{figure}

A natural question arising from this peculiar behavior in the electronic
structure of a GS, which is topologically different from that of
pristine graphene, is how the LLs are distributed.
Figure~\ref{Fig3} shows the calculated LLs of the 1D Kronig-Penney GSs
depicted in Fig.~\ref{Fig1} for various values of $U_0$~\cite{note:Landau_bands}.
When the superlattice potential modulation is moderate [Fig.~\ref{Fig3}(b)],
the spacings between neighboring LLs become smaller than
those in pristine graphene [Fig.~\ref{Fig3}(a)], owing to a reduction in the band velocity.
Once $U_0$ becomes larger than $4\pi\,\hbar v_0/L$ (= 0.4~eV for $L=20$~nm),
the zero-energy LLs
(corresponding to zero carrier density) become three-fold degenerate [Fig.~\ref{Fig3}(d)].
An important point to note is that this degeneracy is insensitive to
$U_0$ over a range of $U_0$ near $6\pi\,\hbar v_0/L$ because the topology of
the electron bands does not
change with this variation~\cite{note:imperfection,note:temperature}.
Moreover, even though the massless particles of the different Dirac cones
may have different band velocities,
the degeneracy of the zero-energy LLs is not affected.

\begin{figure}
\includegraphics[width=0.8\columnwidth]{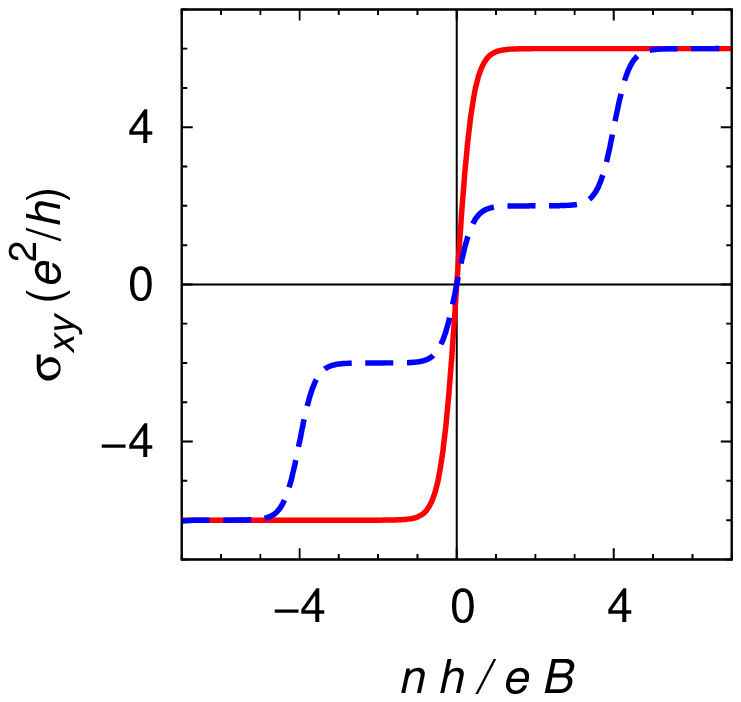}
\caption{(color online) Hall conductivity $\sigma_{xy}$ versus carrier density
(with an artificial broadening for illustration)
for a 1D Kronig-Penney GS with $U_0$ near $6\pi\hbar v_0/L$
(solid red line) is compared to that of pristine graphene
(dashed blue line).}
\label{Fig4}
\end{figure}

The dependence of the Hall conductivity $\sigma_{xy}$ on the charge carrier density $n$
most directly reflects the degeneracy of the LLs.
Figure~\ref{Fig4} schematically shows that,
depending on the superlattice potential parameters,
$\sigma_{xy}$ of the GSs considered has a $4(2N+1)\,e^2/h$ step
as the density is scanned from hole-like to electron-like carriers.
(We have put in the additional factor 4 coming from
the spin and valley degeneracies in this discussion and in Fig.~\ref{Fig4}.)
Because the degeneracy of the LLs in the 1D GSs
is insensitive to a variation in $U_0$,
this qualitative difference in
$\sigma_{xy}$ of the 1D GSs from that of
pristine graphene (Fig.~\ref{Fig4}) is expected to be robust,
and will provide a measurable signature of the unique electronic
structure of the 1D GSs.

In conclusion, we have shown that the electronic structure
of 1D graphene superlattices
can have additional Dirac cones at the
same energy as the original cones at the K and K' points of pristine graphene.
These new massless particles contribute
to a $4(2N+1)$-fold degeneracy in the zero-energy Landau levels,
whose signature is reflected in a $4(2N+1)\,e^2/h$ Hall conductivity step
where $N=0,\,1,\,2,\,...$ depending on the superlattice potential parameters.
This feature of the electronic structure of the 1D graphene superlattices
gives rise to new properties for the quantum Hall effect.
Equally importantly, these new phenomena may provide a direct way to characterize
the peculiar electronic structure of these systems experimentally.

C.\,-H.\,P. thanks Dmitry Novikov and Jay Deep Sau for fruitful discussions.
This work was supported by NSF Grant
No.\,DMR07-05941 and by the Director, Office of Science, Office of Basic Energy
Sciences, Division of Materials Sciences and Engineering Division,
U.S. Department of Energy under Contract No.\,DE- AC02-05CH11231.
Y.\,-W.\,S. was supported by the KRF (KRF 2008-314-C00111)
and Quantum Metamaterials Research Center
(No. R11-2008-053-01002-0) through the KOSEF funded by the MEST.
Computational resources have been provided by NERSC and TeraGrid.

{\it Note added.}--- After submission, we became aware of a recent
theoretical work~\cite{brey_fertig} confirming the newly generated massless
fermions reported in this manuscript, with applications to transport properties.

\section{Supplementary Information}
\subsection{{\large 1. Sinusoidal superlattice}}

\begin{figure*}
\includegraphics[width=1.6\columnwidth]{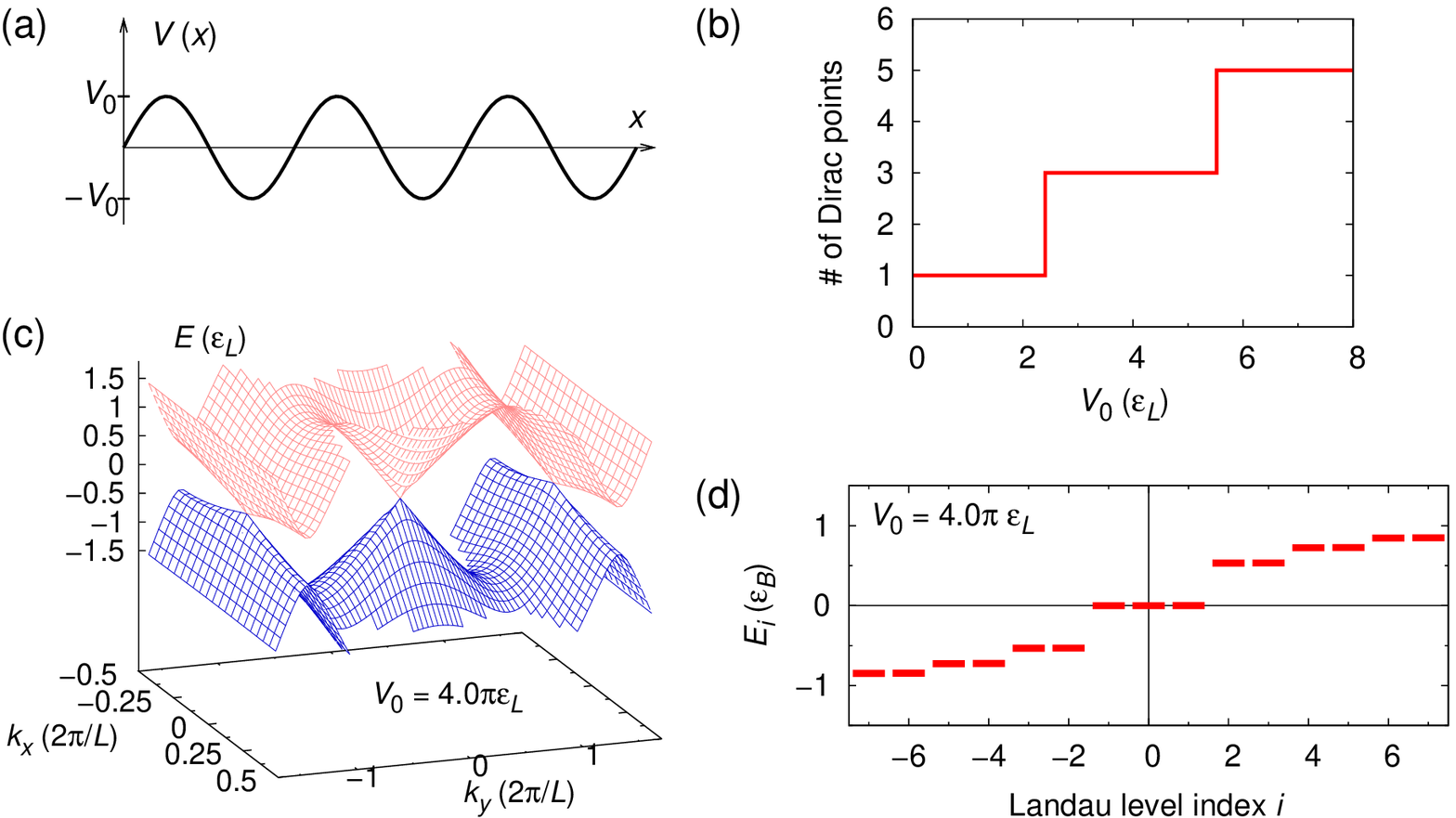}
\caption{(color online) (a) Schematic diagram of a sinusoidal type of potential
applied to graphene with lattice period {\it L} and potential amplitude $V_0$
[$V(x)=V_0\sin(2\pi x/L)$].
(b) Number of Dirac points (not including the spin and valley degeneracies)
in a GS versus $V_0$ in units of $\varepsilon_L$ ($\equiv\hbar v_0/L$;
for example, if $L=20$~nm, $\varepsilon_L=33$~meV).
(c) Electron energy versus wavevector near the original Dirac point
($k_x=k_y=0$) for a GS with $V_0=4.0\pi\varepsilon_L$.
(d) Landau level energy $E_i$ (in units of
$\varepsilon_B\equiv\hbar v_0/l_B$ with $l_B=\sqrt{\hbar c/eB}$)
versus the Landau level index $i$ ($i=0,\,\pm1,\,\pm2,\,...\,$)
in a GS formed with a sinusoidal potential
with $V_0=4.0\pi\varepsilon_L$ and $L=0.5l_B$.
Note the 3-fold degeneracy (becoming 12-fold degeneracy when the spin and
valley degeneracies are considered) around $E_i=0$.}
\label{SFig1}
\end{figure*}

In the main manuscript, we state that the 
essential features in the electronic structure of graphene
superlattices (GSs) revealed by considering the Kronig-Penney
type of potential remain valid for GSs made with different types
of periodic potentials.
In this section, we support this by showing the results
for GSs with sinusoidal types of external periodic potentials [Fig.~\ref{SFig1}(a)].

The function $\alpha(x)$ defined by Eq.~(6) in Ref.~\onlinecite{park:126804}
for a sinusoidal type of external periodic potential
$V(x)=V_0\sin(2\pi x/L)$ is $\alpha(x)=-{V_0 L}/{\pi}\hbar v_0\cdot \cos(2\pi x/L)$.
Therefore, as shown in Eqs.~(9) and~(15) of Ref.~\onlinecite{park:126804},
the group velocity at the original Dirac point
perpendicular to the periodic direction is given by $v_y=f_0 v_0$,
where $v_0$ is the group velocity in pristine graphene and
$f_0=J_0(LV_0/\pi\hbar v_0)$.
Here, $J_0(x)$ is the zeroth order Bessel function of the first kind.
Our calculations show that a new pair of
massless Dirac points are generated whenever $f_0=0$, i.e.,
$V_0$ is equal to
\begin{equation}
V^N_0=\pi x_{0,\,N}\,\frac{\hbar v_0}{L}\,,
\label{eq:VN}
\end{equation}
where $x_{0,\,N}$ is the {\it N}-th root of $J_0(x)$ (e.g., $x_{0,1}=2.405$, $x_{0,2}=5.520$, etc.)
[Fig.~\ref{SFig1}(b)].

Figure~\ref{SFig1}(c) shows the energy bandstructure of a sinusoidal type of GS with
$V_0=4.0\pi\cdot\frac{\hbar v_0}{L}$. Because this value of $V_0$ is between
$V^1_0$ and $V^2_0$,
a pair of new zero-energy massless Dirac cones are generated, and they
clearly affect the Landau level degeneracy [Fig.~\ref{SFig1}(d)] in the same way
as discussed in the main manuscript for a Kronig-Penney type of GS.

\subsection{\large 2. Effects of symmetry breaking on the newly generated massless fermions}

\begin{figure*}
\includegraphics[width=2.0\columnwidth]{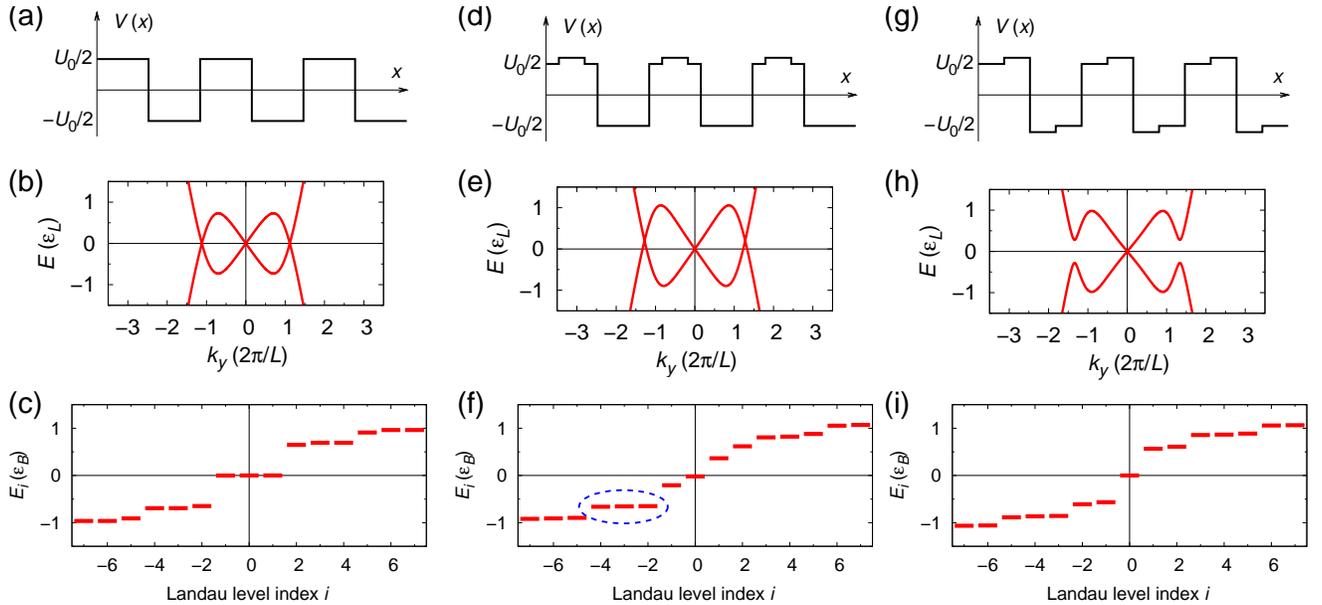}
\caption{(color online) (a) Kronnig-Penney type of potential $V(x)$
given by $U_0/2$ for $0<x<L/2$ and $-U_0/2$ for $L/2<x<L$ with lattice period $L$.
(b)Electron energy (in units of $\varepsilon_L=\hbar v_0/L$)
versus $k_y$ with $k_x=0$ in a GS formed by the periodic potential in (a)
with $U_0=6\pi\varepsilon_L$
(c) Landau level energy $E_i$ (in units of
$\varepsilon_B\equiv\hbar v_0/l_B$ with $l_B=\sqrt{\hbar c/eB}$)
versus the Landau level index $i$ ($i=0,\,\pm1,\,\pm2,\,...\,$)
in a GS depicted in (b), with lattice period $L=0.5l_B$.
(d) to (f): Same quantities as in (a) to (c) for a periodic potential $V(x)$
with a perturbation that breaks the odd symmetry. The perturbing potential
$\Delta V(x)$ within one unit cell is given by $+10~\%$ of the potential
amplitude ($U_0/2$) for $L/8<x<3L/8$
and zero otherwise. Blue dashed circle in (f) shows a three-fold degenerate
set of Landau levels.
(g) to (i): Same quantities as in (a) to (c) for a periodic potential $V(x)$
with a perturbation that breaks the even symmetry. The perturbing potential
$\Delta V(x)$ within one unit cell is given by $+10~\%$ and $-10~\%$ of the potential
amplitude ($U_0/2$) for $L/4<x<L/2$ and for $L/2<x<3L/4$, respectively,
and zero otherwise.}
\label{SFig2}
\end{figure*}

In this section, we discuss the effect of symmetry breaking
of the external periodic potential on the newly generated massless fermions.
(The case of random perturbation is discussed in Ref. 42 of the main manuscript.
In this section, we focus on a periodic perturbing potential that
breaks the even or odd symmetry.)

Figure~\ref{SFig2}(a)-(c) repeats the results shown in the main manuscript for
a Kronig-Penney type of periodic potential having both even and odd symmetries.
If the odd symmetry is broken by adding an appropriate perturbation [Fig.~\ref{SFig2}(d)],
new branches of massless fermions are still generated [Fig.~\ref{SFig2}(e)].
The energy at these new massless
Dirac points however is different from that of the original Dirac point [Fig.~\ref{SFig2}(e)].
Even though $E_i=0$ Landau level does not have the degeneracy coming from
multiple Dirac points, some lower-index Landau levels still show
this kind of degeneracy [Fig.~\ref{SFig2}(f)].

If the even symmetry is broken through a perturbing potential [Fig.~\ref{SFig2}(g)],
new Dirac points are not generated [Fig.~\ref{SFig2}(h)].
However, the signature of newly generated states may still be probed with
photoemission experiments or transport measurements~\cite{brey_fertig}.

\subsection{\large 3. Pseudospins of new massless fermions}

\begin{figure}
\includegraphics[width=1.0\columnwidth]{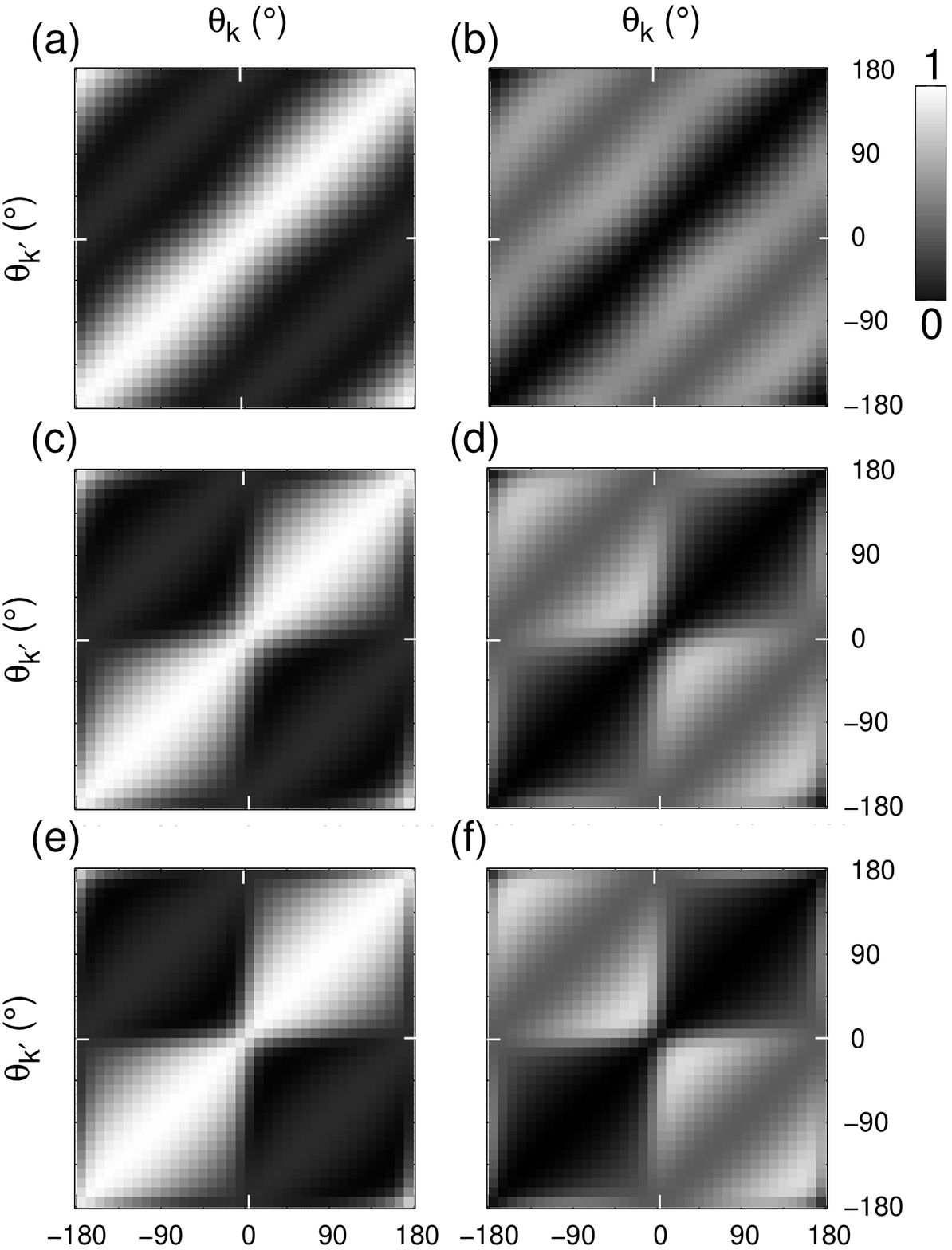}
\caption{(a) and (b): Calculated overlap
of two quasiparticle states $\psi^0_{s,{\bf k}}({\bf r})$ and $\psi^0_{s',{\bf k'}}({\bf r})$,
$\left|\left<\psi^0_{s',{\bf k'}}|e^{i({\bf k'}-{\bf k})\cdot{\bf r}}|\psi^0_{s,{\bf k}}\right>\right|^2$,
in a GS depicted in Fig.~\ref{SFig1}(a)
with $U_0=6\pi\varepsilon_L$
versus $\theta_{\bf k}$ and $\theta_{\bf k'}$ which are the angles between
the $k_x$ axis and wavevectors ${\bf k}$ and ${\bf k'}$,
measured from the newly-generated
massless Dirac point (appearing when $U_0=4\pi\varepsilon_L$ and moving along the $k_y$ direction
as $U_0$ is increased further), respectively.
The overlap is shown in a gray scale (0 in black and 1 in white).
The results show negligible dependence on
$|{\bf k}|$ and $|{\bf k'}|$ when they are smaller than $\sim0.05\times2\pi/L$
(in the figures, $|{\bf k}|=|{\bf k'}|=0.02\times2\pi/L$).
The two states are in the same band ($s'=s$) in (a)
and are in different bands ($s'=-s$) in (b).
(c) and (d), and, (e) and (f): Same quantities as in (a) and  (b) for
GSs with $U_0=8\pi\varepsilon_L$ and $U_0=10\pi\varepsilon_L$,
respectively.}
\label{SFig3}
\end{figure}

The pseudospin character of the newly-generated massless states are different
from that of the original Dirac fermions.
In order to illustrate the pseudospin character of these states,
we numerically calculate the overlap
$\left|\left<\psi_{s',{\bf k}'}|e^{i({\bf k}'-{\bf k})\cdot{\bf r}}|\psi_{s,{\bf k}}\right>\right|^2$
of two quasiparticle states $\psi_{s,{\bf k}}({\bf r})$ and $\psi_{s',{\bf k}'}({\bf r})$
in a GS having wavevectors ${\bf k}$ and ${\bf k}'$ measured from the newly-generated
Dirac point (appearing when $U_0=4\pi\varepsilon_L$ and moving along the $k_y$ direction
as $U_0$ is increased further).
The behavior shown in Fig.~\ref{SFig3}, which corresponds to the overlap of the
pseudospin part of the wavefunctions, is robust if the magnitudes of ${\bf k}$ and ${\bf k}'$
are smaller than $\sim0.05\times2\pi/L$.

As mentioned in the main manuscript, the pseudospin character of these additional
massless fermions [e.\,g.\,, the left and the right Dirac cones (not the center one)
in Fig.~\ref{Fig1}(c)] are different from that of the original
massless Dirac fermions. Backscattering amplitude due to a slowly varying potential
within one of the new cones does not vanish (Fig.~\ref{SFig3}).


\begin{thebibliography}{10}

\bibitem{berger:2006Graphene_epitaxial}
C. Berger {\it et~al.}, Science {\bf 312},  1191  (2006).

\bibitem{novoselov:2005Nat_Graphene_QHE}
K.~S. Novoselov {\it et~al.}, Nature {\bf 438},  197  (2005).

\bibitem{zhang:2005Nat_Graphene_QHE}
Y. Zhang, J.~W. Tan, H.~L. Stormer, and P. Kim, Nature {\bf 438},  201  (2005).

\bibitem{wu:026801}
X. Wu {\it et~al.}, Phys. Rev. Lett. {\bf 101},  026801  (2008).

\bibitem{katsnelson:2006NatPhys_Graphene_Klein}
M.~I. Katsnelson, K.~S. Novoselov, and A.~K. Geim, Nature Phys. {\bf 2},  620
  (2006).

\bibitem{stander:026807}
N. Stander, B. Huard, and D. Goldhaber-Gordon, Phys. Rev. Lett. {\bf 102},
  026807  (2009).

\bibitem{young:2009NatPhys}
A.~F. Young and P. Kim, Nature Phys. {\bf 5},  222  (2009).

\bibitem{bai:2007PRB_Graphene_SL}
C. Bai and X. Zhang, Phys. Rev. B {\bf 76},  075430  (2007).

\bibitem{park:2008NatPhys_GSL}
C.-H. Park {\it et~al.}, Nature Phys. {\bf 4},  213  (2008).

\bibitem{barbier:115446}
M. Barbier, F.~M. Peeters, P. Vasilopoulos, and J. J.~Milton~Pereira, Phys.
  Rev. B {\bf 77},  115446  (2008).

\bibitem{park:2008NL_Supercollimation}
C.-H. Park {\it et~al.}, Nano Lett. {\bf 8},  2920  (2008).

\bibitem{park:126804}
C.-H. Park {\it et~al.}, Phys. Rev. Lett. {\bf 101},  126804  (2008).

\bibitem{masir:235443}
M.~R. Masir, P. Vasilopoulos, A. Matulis, and F.~M. Peeters, Phys. Rev. B {\bf
  77},  235443  (2008).

\bibitem{masir:035409}
M.~R. Masir, P. Vasilopoulos, and F.~M. Peeters, Phys. Rev. B {\bf 79},  035409
   (2009).

\bibitem{dellanna:045420}
L. Dell'Anna and A.~D. Martino, Phys. Rev. B {\bf 79},  045420  (2009).

\bibitem{ghosh:arxiv}
S. Ghosh and M. Sharma, J. Phys. Cond. Matt. {\bf 21},  292204  (2009).

\bibitem{guinea:075422}
F. Guinea, M.~I. Katsnelson, and M.~A.~H. Vozmediano, Phys. Rev. B {\bf 77},
  075422  (2008).

\bibitem{wehling:EPL2008_graphene_corrugation}
T.~O. Wehling, A.~V. Balatsky, M.~I. Katsnelson, and A.~I. Lichtenstein,
  Europhys. Lett. {\bf 84},  17003  (2008).

\bibitem{isacsson:035423}
A. Isacsson, L.~M. Jonsson, J.~M. Kinaret, and M. Jonson, Phys. Rev. B {\bf
  77},  035423  (2008).

\bibitem{meyer:123110}
J.~C. Meyer, C.~O. Girit, M.~F. Crommie, and A. Zettl, Appl. Phys. Lett. {\bf
  92},  123110  (2008).

\bibitem{marchini:2007PRB_Graphene_Ru}
S. Marchini, S. G\"{u}nther, and J. Wintterlin, Phys. Rev. B {\bf 76},  075429
  (2007).

\bibitem{vazquez:2008PRL_Graphene_SL}
A.~L. Vazquez~de Parga {\it et~al.}, Phys. Rev. Lett. {\bf 100},  056807
  (2008).

\bibitem{sutter:2008NatMat}
P.~W. Sutter, J.-I. Flege, and E.~A. Sutter, Nature Mater. {\bf 7},  406
  (2008).

\bibitem{martoccia:126102}
D. Martoccia {\it et~al.}, Phys. Rev. Lett. {\bf 101},  126102  (2008).

\bibitem{pan:2007condmat_Graphene_SL}
Y. Pan {\it et~al.}, arXiv:0709.2858v1.

\bibitem{coraux:2008NL}
J. Coraux, A.~T. N'Diaye, C. Busse, and T. Michely, Nano Lett. {\bf 8},  565
  (2008).

\bibitem{ndiaye:2008NJP}
A.~T. N'Diaye1 {\it et~al.}, New J. Phys. {\bf 10},  043033  (2008).

\bibitem{pletikosic:056808}
I. Pletikosi\'{c} {\it et~al.}, Phys. Rev. Lett. {\bf 102},  056808  (2009).

\bibitem{PhysRevLett.95.146801}
V.~P. Gusynin and S.~G. Sharapov, Phys. Rev. Lett. {\bf 95},  146801  (2005).

\bibitem{mccann:086805}
E. McCann and V.~I. Fal'ko, Phys. Rev. Lett. {\bf 96},  086805  (2006).

\bibitem{novoselov:2006NatPhys}
K.~S. Novoselov {\it et~al.}, Nature Phys. {\bf 2},  177  (2006).

\bibitem{note:degeneracy}
The zigzag form of the vector potential we used results in the perpendicular
  magnetic field of a uniform strength pointing along the $+z$ direction for
  half the artificial periodicity and along the $-z$ direction for the other
  half.

\bibitem{note:diff_new_gen}
Note that these additional massless fermions with $k_x=0$ are different from
  those generated at the supercell Brillouin zone boundaries discussed in
  Ref.~\onlinecite{park:126804}.

\bibitem{note:vperp}
The group velocity along the $k_y$ direction is given by
  $v_y=v_0\,\int_{-L/2}^{L/2}e^{i\alpha(x)}\,dx$ where $\alpha(x)$ for a
  Kronig-Penney type of superlattice is given by $\alpha(x)=U_0/\hbar
  v_0\cdot\left(|x|-L/4\right)$ (see Ref.~\onlinecite{park:126804}). When
  Eq.~(\ref{eq:U0}) is satisfied, $v_y=0$.

\bibitem{note:validity_Dirac}
{The use of the Dirac equation for this problem is still valid because the new
  Dirac points are very close to the original Dirac point ($k_x=k_y=0$), inside
  the regime where the graphene band is linear. For example, in Fig.~1(c), the
  new massless Dirac points appear at $k_y=1.1\cdot2\pi/L=0.034$~\AA\ if
  $L=20$~nm}.

\bibitem{park:unp_chiral}
See Supplementary Information.

\bibitem{PhysRevLett.62.1177}
R.~W. Winkler, J.~P. Kotthaus, and K. Ploog, Phys. Rev. Lett. {\bf 62},  1177
  (1989).

\bibitem{note:Landau_bands}
{When a vector potential ${\bf A}(x)=B\,x\,{\hat {y}}$ is used, the shift in
  the center of a Landau state along the {\it x} direction is proportional to
  $k_y$; hence, in a 1D superlattice periodic along {\it x}, the external
  periodic potential felt by a Landau state varies with $k_y$, resulting in a
  finite Landau band width $\Delta E$ (Ref.~\onlinecite{PhysRevLett.62.1177}).
  However, if the level spacing between LLs is much larger than $\Delta E$, the
  signature of these LLs can be measured from experiments, as in graphene
  (Refs.~\onlinecite{park:massless2DEG} and~\onlinecite{gibertini:241406}). In
  our case, we have checked that for the zero-energy LLs plotted in
  Fig.~\ref{Fig3}, as long as $l_B>L$, $\Delta E$ is smaller than 0.2~\% of
  $\varepsilon_B$. Thus, in the conditions considered here, Landau bands can be
  considered as discrete levels, i.\,e.\,, LLs}.

\bibitem{note:imperfection}
The effect of up to~$\sim$10~\% disorder in the on-site potential of graphene
  on the LL broadening is small (Ref.~\onlinecite{zhu:056803}). Since the
  low-energy electronic band structure of a GS can effectively be decomposed
  into a finite number of Dirac cones, we expect that the LLs in GSs with
  moderate disorder can be observed.

\bibitem{note:temperature}
For a Kronig-Penney superlattice with $L=20$~nm and $l_B=2L$ (corresponding to
  $B=0.4$~T), the magnetic energy $\varepsilon_B$ is 17~meV. In this case,
  temperature has to be lowered to observe the LL quantization.

\bibitem{brey_fertig}
L. Brey and H.~A. Fertig, Phys. Rev. Lett. {\bf 103}, 46809 (2009).

\bibitem{park:massless2DEG}
C.-H. Park and S.~G. Louie, Nano Lett. {\bf 9},  1793  (2009).

\bibitem{gibertini:241406}
M. Gibertini {\it et~al.}, Phys. Rev. B {\bf 79},  241406  (2009).

\bibitem{zhu:056803}
W. Zhu {\it et~al.}, Phys. Rev. Lett. {\bf 102},  056803  (2009).

\end{thebibliography}
\end{document}